\documentclass[aps,showpacs,epsf,preprint]{revtex4}
\usepackage{graphicx}
\usepackage{dcolumn}
\usepackage{amsmath}
\usepackage{amssymb}
\usepackage{empheq}
\usepackage[colorlinks]{hyperref}
\hypersetup{citecolor=blue}
\usepackage{bm}
\usepackage{epstopdf}
\usepackage{amsthm}
\usepackage{float}
\graphicspath{{./figures/}}
\usepackage{accents}
\usepackage{comment}

\def\A{\alpha}

\def\epsilon{\varepsilon}

\def\t{\tau}

\def\epsilon{\varepsilon}

\def\K{{\cal K}}
\def\a{{\cal A}}
\def\A{\mathbf{A}}
\begin{document}


\title{Phase-locking in $k$-partite networks of delay-coupled oscillators}

\author{Joydeep Singha}
\email{joydeepsingha105@gmail.com}
\affiliation{Department of Chemistry, Indian Institute of Technology Delhi, New Delhi - 110016}
\author{Ramakrishna Ramaswamy}
\email{ramaswamy@iitd.ac.in}
\affiliation{Department of Chemistry, Indian Institute of Technology Delhi, New Delhi - 110016}
\pacs{05.45.Ra, 05.45.-a, 05.45.Df, 64.60.Ak}
\date{\today}

\begin{abstract}
We examine the dynamics of an ensemble of phase oscillators that are divided in $k$ sets, with time-delayed coupling interactions {\em only} between oscillators in different sets or partitions. The network of interactions  thus forms a $k-$partite graph. A variety of phase-locked states are observed; these include, in addition to the fully synchronized in-phase solution, splay cluster solutions in which all oscillators within a  partition are synchronised and the phase differences between oscillators in different partitions are multiples of $2\pi/k$. Such solutions exist independent of the delay and we determine the generalised stability criteria for the existence of these phase-locked solutions  for the $k-$partite system. With increase in time-delay, there is an increase in multistability, the generic solutions coexisting with a number of other partially synchronized solutions. The Ott-Antonsen ansatz is applied for the special case of a symmetric $k-$partite graph to obtain a single time-delayed differential equation for the attracting synchronization manifold. Agreement with numerical results for the specific case of oscillators on a tripartite lattice (the $k=3$ case) is excellent.  
\end{abstract}

\maketitle

\section{\label{sec_1}Introduction}
\par 
The synchronisation of ensembles of coupled dynamical systems with specific structural features is a subject of continuing interest, particularly in the context of the emergence of complex dynamical states. Over the past decades studies have explored a variety of systems on networks of varying topology and have helped clarify the role of the topology of the underlying connectivity and the resulting dynamics \cite{strogatz2001, albert2002, mendes2002, newman2003, jeong2000, newman2001}. Graphs with specific substructures such as communities or partitions have been studied quite extensively, the bipartite case, namely with two partitions, being one of the most common. Indeed, simple networks such as the linear chain or star can be viewed as being bipartite \cite{jeong2000, newman2001, tanaka2005, saavedra2009} since the nodes on the network can be divided into two groups, and links can only exist between nodes in different groups. 

\par
\begin{figure}
\includegraphics[scale = 0.4]{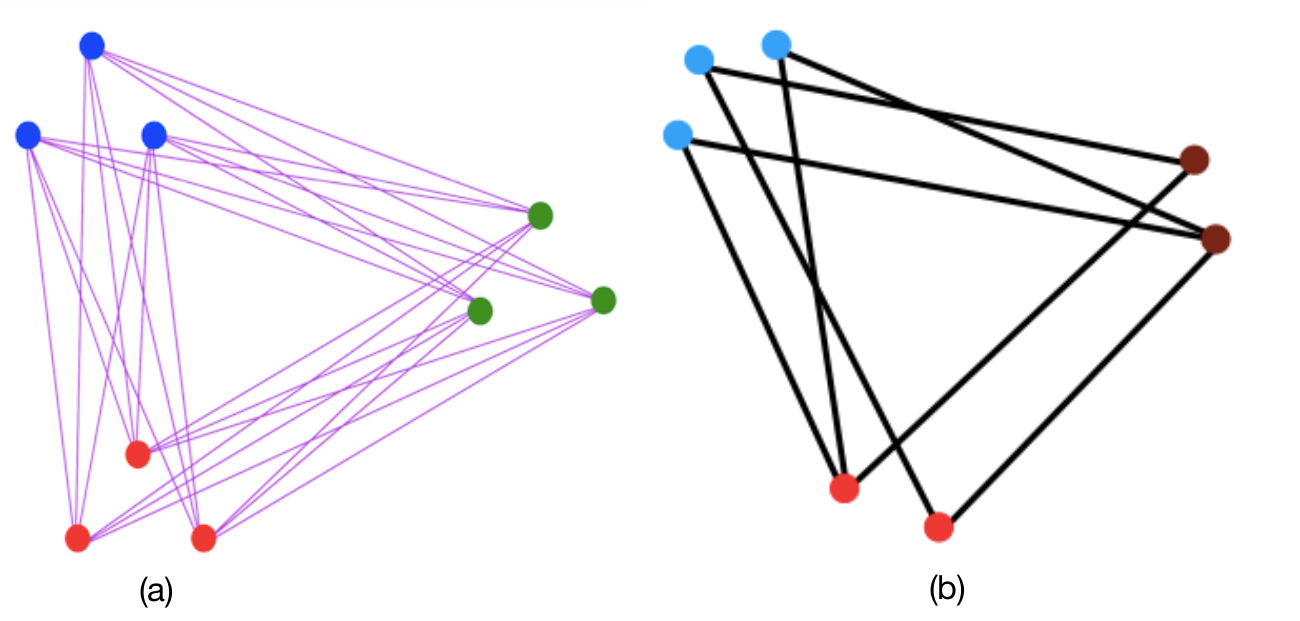}
\caption{\footnotesize Examples of tripartite networks. Nodes in the different partitions have different colours, and links are forbidden within a partition. Shown in a) is the complete 3-partite graph, denoted $K_{3,3,3}$ with three nodes in each partition and with the connection being all-to-all, namely every node is connected to every other allowed node. In b) there are unequal numbers of nodes in the partitions, and the connections are also asymmetric.}
\label{fig: scheme}
\end{figure}

  In this paper we study the dynamics of oscillators on multipartite networks \cite{agnar2007}, namely the dynamics of $k$ sets of oscillators organized such that there are no interactions among the elements of a given set or partition, and with coupling only between elements of different partitions. (Examples of the $k$ = 3 case or tripartite networks are shown in Fig. \ref{fig: scheme}.) Our interest is in the nature of the collective dynamics that results as a consequence of the underlying multipartite network topology. Similar questions have been explored extensively in recent years in the context of complex networks \cite{barbillon2020, shai2017, namgil2021, maslov2002, white2004, fredrik2003, katy2004, lu2012, antonia2016}, and while the case of partitioned networks with $k$=2 has been studied in some detail, the case of general $k$ has not been examined in relation to the dynamics. Accordingly, here we study the simplest example, a set of  coupled phase oscillators on multipartite graphs with time-delay coupling (for increased generality). In addition to the globally synchronized state that can be anticipated to arise in such systems for sufficiently strong coupling, one can expect that there will be a variety of phase-locked solutions. 

\par Multipartite networks are relevant in numerous contexts and have recently attracted considerable attention. Tripartite networks are useful in understanding the relationship of plants, pollinators and herbivores in ecology, for instance, or between crop species, seeds and farmers in ethnobiology  \cite{barbillon2020}. The structure and dynamics of multilayer ecological networks between species \cite{shai2017} can also be seen as a multipartite network, and other situations where this is useful is in the study of medicinal plants, their chemical components and the physiological targets \cite{namgil2021}. Metabolic signalling pathways constructed from protein and gene regulatory networks \cite{maslov2002} are multipartite, as are kinship and marriage networks \cite{white2004} that can be used to understand social ties and integration. Other contexts where these networks have been used are in  the transmission of sexual disease \cite{fredrik2003}, collaboration networks between authors, articles and research fields \cite{katy2004}, product recommendation systems as well as modelling and prediction of user demands \cite{lu2012, antonia2016}.

\par Dynamics on bipartite graphs have been studied earlier, starting with the work of Schuster and Wagner \cite{schuster1989} who examined the case of two delay-coupled Kuramoto oscillators and described the solutions in detail. They also discussed the nature of multistability as well as the basins of attraction of coexisting phase locked solutions. The role of delayed feedback on the transition to synchrony was studied by Goldobin and Pikovsky \cite{goldobin2006} for a globally coupled ensemble of Kuramoto oscillators. Earl and Strogatz \cite{earl2003} investigated the stability of the phase locked solutions in the general case of delay-coupling and provided a stability criterion for the in-phase synchronised solutions. Punetha et.al. \cite{punetha2015} showed that a discrete time-delay, can lead to remote synchronisation of the oscillators for appropriate frequency mismatch between the partitions. For a general bipartite system of oscillators, the exact stability criteria for in-phase and anti-phase solutions have been obtained \cite{atay2015}, and it has been proved that these are indeed the generic solutions. 

\par The present analysis proceeds along similar lines. We show that a generalised $k-$partite system of identical oscillators generically supports several phase-locked solutions. In addition to the completely synchronised state where all phases are identical, there are splay solutions with all oscillators in a given partition having identical phases,  with phase differences that are multiples of $2\pi/k$ between partitions: this is a generalisation of the antiphase solution when there are only two partitions.  An explicit proof of the generalised stability criteria for these as well as other solutions that can arise for a given value of $k$ is provided. This is applicable both to the case of identical oscillators and non-identical oscillators, and for a class of smooth coupling functions. The generalised $k-$partite system supports multiple stable dynamical states, and their number (which can be estimated analytically) increases linearly with the delay parameter. Utilising the Ott-Antonsen ansatz \cite{OA_1, OA_2, OA_3}, we find that under special conditions the system can be reduced to a single delay differential equation. 
 
\par This paper is organised in a fairly linear fashion. In Section \ref{sec_2} we describe the specific dynamical system studied and find the generic phase locked solutions, namely the in-phase and splay-phase clusters. A specific case of the tripartite system is considered and its phase locked solutions are found in the section \ref{sec_2a}. Section \ref{sec_2b} discusses the nature of the multistability of the solutions in the $k-$partite system. This is followed by a discussion of  the Ott--Antonsen ansatz and the subsequent dimensionality reduction of the multipartite system in Section \ref{sec_3}. In both Sections \ref{sec_2} and \ref{sec_3}, numerical results are presented for a tripartite system of oscillators for the cases when all frequencies are identical as well as when frequencies in different partitions are distinct, in order to compare with the corresponding analytical results.  We conclude in Section \ref{sec_4} with a summary.

\section{\label{sec_2}Phase oscillators on a $k-$partite network}
\par Consider a lattice with partitions numbered $j = 1, 2, \ldots k$, with $N_j$ nodes in partition $j$. The edges connecting the nodes are undirected, and the number of nodes in the various partitions are unequal in general. The underlying network therefore need not be a complete $k-$partite graph although we require that every node be connected to at least one node in each of the other partitions.  

The total number of nodes is $N=\sum_jN_j$. At each node there is a phase oscillator,  with oscillator variable denoted $\theta^{j}_{i}$:  the superscript $j = 1,2,\ldots, k$ labels the partition and the subscript $i=1,2,\ldots, N_j$ indexes the nodes within the partition.  All oscillators within a partition are taken to have the same frequency, so in the absence of coupling a given oscillator evolves according to the dynamics
\begin{equation}
\dot{\theta}^{j}_{i}(t) = \omega_{j}.
\end{equation}
In the coupled system, the dynamics is governed by the evolution equation
\begin{equation}
\dot{\theta}^{j}_{i}(t) = \omega_{j} + \epsilon \sum_{m=1}^{k}\sum_{\ell=1}^{N_m}\frac{\a^{jm}_{i\ell}}{K^{jm}_{i}}\sin(\theta^{m}_{\ell}(t - \tau) - \theta^{j}_{i}(t)),
\label{Eq: evolve}
\end{equation}
where $j, m = 1, 2, \cdots, k$, $j \neq m$ and $\epsilon$ is the coupling strength and the parameter $\tau$ represents the transmission delay between the partitions. The adjacency matrices $\A^{jm}$ have dimension $N_j\times N_m$ and are in general non-square and asymmetric. Since we consider undirected edges, the matrix elements $\a^{jm}_{i\ell}$ are equal to one if node $i$ in partition $j$ is connected to node $\ell$ in partition $m$, and zero otherwise. The normalising factor is defined as
\begin{equation}
    K^{jm}_{i} = \sum\limits_{\ell = 1}^{N_m}\a^{jm}_{i\ell}    
\end{equation}
and the degree of a given node $i$ in partition $j$ is $\K^{j}_{i} = \sum\limits_{m = 1}^{k}K^{jm}_{i}$. As mentioned above, we require that every node has at least one link to a node in each of the other partitions, and therefore $K^{jm}_{i} > 0$. The adjacency matrix of the entire lattice $\A$ is composed of the blocks, $\A^{jm}$. For a tripartite system, for example, the adjacency matrix is  
\begin{equation}
\A = \begin{bmatrix}
    		\A^{11} & \A^{12} & \A^{13}\\
         	\A^{21} & \A^{22} & \A^{23}\\
         	\A^{31} & \A^{32} & \A^{33}\\
      	\end{bmatrix}.
        \label{Eq: adj_mat}
\end{equation}
All entries of $\A^{ii}, i=1,2,3$  are zero, and $\A^{jm} = (\A^{mj})^{\intercal}$. Clearly $\A$ has dimension $N\times N$. 

\par In the globally synchronised state, all the oscillators are locked to a common frequency which we denote by $\Omega$. Consider a phase-locked solution where for all oscillators $i$ in partition $j$ the phase is 
\begin{equation}
            \theta^{j}_{i} = \Omega t - \mu_j,
    \label{Eq: gen_soln}
\end{equation}
with $\mu_1$ chosen to be zero.  Oscillators in each partition are in synchrony and the different partitions are phase--shifted from each other.  Substituting this solution into the evolution equation (\ref{Eq: evolve}) we get,
\begin{equation}
        \Omega = \omega_j - \sum\limits_{m = 1, m \neq j}^{k}\epsilon\sin(\Omega\tau +\mu_m - \mu_j).
        \label{Eq: OMG}
    \end{equation}
The above set of $k$ equations are always satisfied by the phase locked solutions in Eq.~(\ref{Eq: gen_soln}). When all intrinsic frequencies are identical, $\omega_{j}= \omega$, there is the in--phase synchronized solution with all  $\mu_j$ = 0 and with average frequency $\Omega_0$ given by the solution to the transcendental equation
\begin{equation}
    \Omega_0 = \omega - (k - 1)\epsilon\sin(\Omega_0\tau).
    \label{Eq: trans_1}
\end{equation}
There can also be a $k-$cluster solution, namely $\mu_j = {2\pi(j - 1)}/{k}, j = 1, \ldots, k$ and satisfies the transcendental equation,
\begin{align}
\begin{split}
    \Omega_c &= \omega - \sum\limits_{j = 1}^{k - 1}\epsilon\sin\left(\Omega_c\tau + \dfrac{2\pi j}{k}\right).
    \label{Eq: trans_2}
\end{split}
\end{align}
A solution of this type is termed a splay phase and known to occur in a variety of dynamical systems; this is the generalisation of the anti-phase solution that occurs in  the bipartite system \cite{punetha2015, atay2015}. Rewriting Eq.~(\ref{Eq: trans_2}) as 
\begin{equation}
\Omega_c = \omega - \epsilon\operatorname{Im}\left[ e^{\iota \Omega_c \tau}\sum\limits_{j = 1}^{k - 1} e^{\frac{\iota 2\pi j}{k}}\right]
\label{Eq: splay_sum}
\end{equation}
and summing the geometric series, it is easy to see that the second term on the right is 
\begin{equation}
-\epsilon\operatorname{Im}\left[ e^{\iota\Omega_c\tau} e^{\frac{\iota2\pi}{k}}\frac{1 - e^{\frac{\iota2\pi(k - 1)}{k}}}{1 - e^{\frac{\iota2\pi}{k}}}\right],
\end{equation}
which can be further simplified to 
\begin{equation}
-\epsilon\operatorname{Im}\left[ e^{\iota\Omega_c\tau}e^{\frac{\iota2\pi}{k}}\frac{e^{\frac{\iota\pi(k - 1)}{k}}}{e^{\frac{\iota\pi}{k}}}	\frac{\sin\frac{\pi(k - 1)}{k}}{\sin\frac{\pi}{k}}\right].
\end{equation}
Now since $e^{\frac{\iota2\pi}{k}}e^{\frac{\iota\pi(k - 1)}{k}} = -e^{\frac{\iota\pi}{k}}$, and $\sin\left( \frac{\pi(k - 1)}{k}\right) = \sin(\frac{\pi}{k})$ 
Eq.~(\ref{Eq: splay_sum}) reduces to
\begin{equation}
\Omega_c = \omega + \epsilon\sin\Omega_c\tau.
\label{Eq: trans_3}
\end{equation}
The common frequency $\Omega_c$ is thus independent of the number of partitions $k$.
Apart from these generic solutions which exist for all values of $\tau$, there can be other phase locked solutions as well, and these can be calculated on a case-by-case basis, depending on the number of partitions in the network and the parameters $\epsilon$ and $\tau$. 

\subsection{\label{sec_2a}The tripartite case, \texorpdfstring{$k$ = 3}{}}
The in--phase and splay solutions can be obtained from Eq.~(\ref{Eq: trans_1}) and (\ref{Eq: trans_2}). However apart from these there are other phase locked solutions that are specific to the value of $k$ under consideration. For the tripartite case $k$ = 3 for example, from the evolution equation we obtain the following three conditions for $\Omega$ via Eq.~(\ref{Eq: OMG}), 
\begin{eqnarray}
\Omega &=& \omega_1 - \epsilon\sin(\Omega\tau + \mu_2) - \epsilon\sin(\Omega\tau + \mu_3)\label{Eq: tri_1},\\
\Omega &=& \omega_2 - \epsilon\sin(\Omega\tau - \mu_2) - \epsilon\sin(\Omega\tau - (\mu_2 - \mu_3))\label{Eq: tri_2},\\
\Omega &=& \omega_3 - \epsilon\sin(\Omega\tau - \mu_3) - \epsilon\sin(\Omega\tau + (\mu_2 - \mu_3))\label{Eq: tri_3}.
\end{eqnarray}
When $\omega_1 = \omega_2 = \omega_3$, subtracting Eq.~(\ref{Eq: tri_3}) from Eq.~(\ref{Eq: tri_2}) we obtain
\begin{equation}
-\epsilon\sin(\Omega\tau - \mu_2) + \epsilon\sin(\Omega\tau - \mu_3) - \epsilon\sin(\Omega\tau - \delta) + \epsilon\sin(\Omega\tau + \delta) = 0,
\end{equation}
namely
\begin{equation}
2\epsilon\cos(\Omega\tau - \phi)\sin\frac{\delta}{2} = -2\epsilon\cos(\Omega\tau)\sin\delta = -4\epsilon\cos(\Omega\tau)\sin\frac{\delta}{2} \cos\frac{\delta}{2},
\label{Eq: soln_1}
\end{equation}
or 
\begin{equation}
\sin\frac{\delta}{2}\left( \cos(\Omega\tau - \phi) + 2\cos\Omega\tau\cos\frac{\delta}{2}\right) = 0
\label{Eq: soln_2}
\end{equation}
where $\delta = \mu_2 - \mu_3$, $\phi = (\mu_2 + \mu_3)/2$. The above equation can be solved easily, giving 
\begin{equation}
    \setlength{\arraycolsep}{0pt}
    \delta = \left\{ \begin{array}{l l}
             &{} 2n\pi,\\
             &{} \pm\cos^{-1}\left(-\frac{\cos(\Omega\tau - \phi)}{2\cos\Omega\tau}\right) + 2n\pi,\\
        \end{array} \right.
\label{Eq: delta_soln}
\end{equation}
where $n \in \mathbb{Z}$.Replacing $\delta$ in Eq.~(\ref{Eq: tri_1}) yields the corresponding solutions of $\phi$ as
\begin{equation}
    \setlength{\arraycolsep}{0pt}
    \phi = \left\{ \begin{array}{l l}
        &{} \sin^{-1}\left( \pm\frac{\omega - \Omega}{2\epsilon}\right)  - \Omega\tau,\\
        &{} \frac{1}{2}\cos^{-1}\left[ 2\cos\Omega\tau\left(\frac{\Omega - \omega}{\epsilon}\right) - \sin2\Omega\tau\right].\\
    \end{array} \right.
    \label{Eq: Phi_soln}
\end{equation}
The corresponding locked frequency $\Omega$ is the root of the transcendental equation obtained by summing Eq.~(\ref{Eq: tri_2}) and (\ref{Eq: tri_3}),  
\begin{equation}
    \omega - \epsilon\sin(\Omega \tau - \phi)\cos\delta/2 
    - \epsilon\sin\Omega\tau\cos\delta -\Omega = 0.
    \label{Eq: trans}
\end{equation}

If the intrinsic frequencies of oscillators in only two of the partitions are equal, namely $ \omega_2 = \omega_3 = \omega \neq \omega_1$, a similar calculation for a phase-locked solution gives $\delta$ as in Eq.~(\ref{Eq: delta_soln}) but with a modified average phase $\phi$,
\begin{equation}
    \setlength{\arraycolsep}{0pt}
    \phi = \left\{ \begin{array}{l l}
        &{} \sin^{-1}\left( \pm\frac{\omega_1 - \Omega}{2\epsilon}\right)  - \Omega\tau,\\
        &{} \frac{1}{2}\cos^{-1}\left[ 2\cos\Omega\tau\left(\frac{\Omega - \omega_1}{\epsilon}\right) - \sin2\Omega\tau\right].\\
    \end{array} \right.
    \label{Eq: phi_soln}
\end{equation}

\begin{figure}[H]
    \centering
    \includegraphics[scale = 0.6]{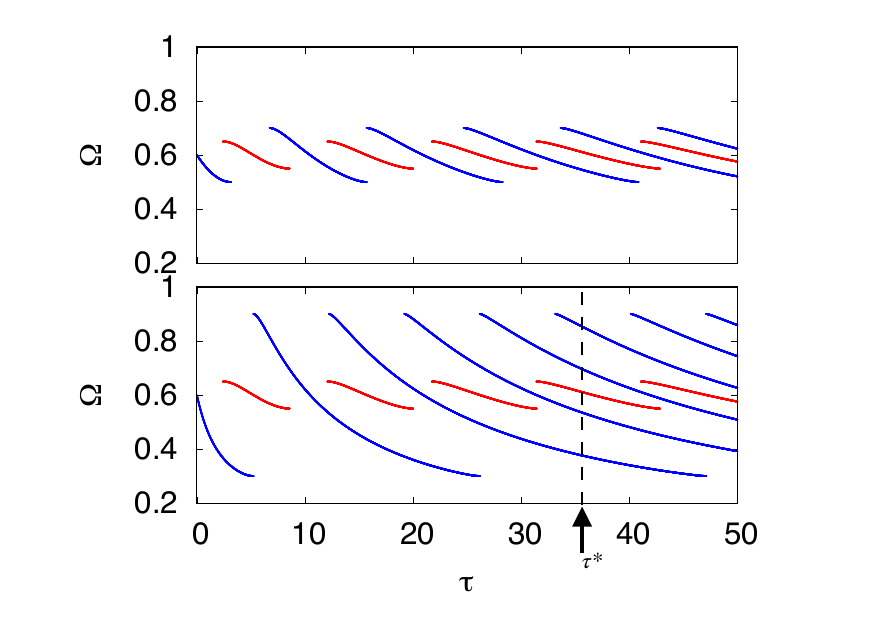}
\caption{\footnotesize The frequency $\Omega$ of the stable branches of generic solutions, {\em i. e.} the in-phase (in blue) and splay phase (in red), shown as a function of the delay parameter $\tau$ for the cases of $k$=3  (above) and $k$=7  (below). The coupling strength is set to $\epsilon = 0.05$ and the intrinsic frequencies $\omega$ of all the oscillators are $0.6$. With increasing numbers of partitions, the range of the branches of the in-phase solutions (blue) increase in both the  $\Omega$ and $\tau$ axes. The upper and lower limits for each branch, respectively increase and decrease in value as more partitions are added (see the discussion in Sec. \ref{sec_2b}). However the stable branches for the splay phase solutions remain unchanged since Eq.~(\ref{Eq: trans_2})  does not depend on $k$. The number of coexisting stable solutions at a given value of $\tau^{*}$ is the number of intersections of a vertical line at $\tau^{*}$ with the branches.}
\label{fig: freq_sync_splay}
\end{figure}

\par The average frequency $\Omega$ can be numerically computed from Eq.~(\ref{Eq: trans}) and the stability criteria for these solutions can be found in a straightforward manner.  Fig.~\ref{fig: freq_sync_splay} shows results using Eqs.~(\ref{Eq: trans_1}) and (\ref{Eq: trans_2}) for the stable in-phase and splay phase cluster solutions. Notice that there is a range of values of $\tau$ where there is only a single stable state, as well as parameter ranges when there are several stable frequencies \cite{schuster1989}. The width of this latter range increases with increasing coupling strength \cite{punetha2015}.

\par The stability criteria for the phase locked solutions can be derived and are given in the Appendix. These are applicable for a class of $C^1$ smooth coupling functions that are $2\pi$-periodic, and the in-phase state is stable if and only if $\epsilon \cos \Omega \tau >0$  and unstable otherwise. In case of nonzero $\mu_{j}$, the solutions are stable iff {\em all} of the following $k$ quantities, 
\begin{equation} 
S_j = \sum\limits_{m = 1, m \neq j}^{k}\epsilon \cos(-\Omega \tau - \mu_m + \mu_j), ~~~~~j = 1, 2, \ldots, k
\label{Eq: splay_stability}
\end{equation}
are positive. When $\mu_j = \frac{2\pi (j - 1)}{k}$, the sum in Eq.~(\ref{Eq: splay_stability}) can be evaluated in a manner similar to that done above for Eq.~(\ref{Eq: splay_sum}), 
giving the stability criteria for splay solutions as, 
\begin{equation}
S_j = \sum\limits_{j = 1}^{k - 1}\epsilon\cos\left(\Omega_c\tau + \frac{2\pi j}{k}\right) = -\epsilon\cos\Omega_c\tau > 0,
\end{equation}
 for all values of $j$. 
For equal intrinsic frequency, Fig. \ref{fig: freq_sync_splay} shows the stable branches of $\Omega_0$ and $\Omega_c$ with respect to the delay parameter $\tau$ for the cases of $k$ = 3 and 7. The widths of the different solution branches and their stability characteristics can be analysed in detail as shown below.

\subsection{\label{sec_2b}Solution branches and multistability}
\begin{figure}[H]
\centering
\includegraphics[scale = 0.7]{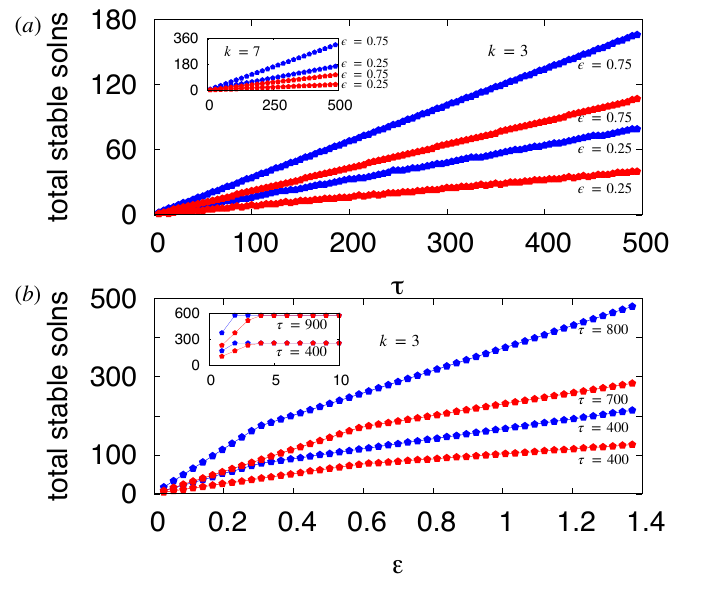}
\caption{\footnotesize The total number of stable in--phase (blue) and splay phase (red) solutions as a function of  (a) the delay $\tau$  and (b) the coupling $\epsilon$. These increase linearly for both types of generic states with $\tau$ for both instances of $k$ (see the inset in (a) as well). Similar behaviour is observed for low $\epsilon$, although as suggested by Eq.~(\ref{overlaps}) in Sec. \ref{sec_3}, the rate of growth will decrease for higher $\epsilon$. The intrinsic frequencies of all the oscillators are 0.6.}
\label{fig: freq_sync_splay_eps}
\end{figure}

The extent of the regions where the generic solutions i.e. in--phase and splay phase locked states are stable, can be determined as a function of $\tau$ and $\epsilon$. The stability conditions of the in--phase solutions, $\epsilon\cos(\Omega_0\tau) > 0$ and the splay solutions, i.e. $\epsilon\cos(\Omega_c\tau) < 0$ give the limitations on $\Omega_0\tau$ and $\Omega_c\tau$ respectively, 
\begin{align}
\begin{split}
-\frac{\pi}{2} + 2n\pi &< \Omega_0\tau < \frac{\pi}{2} + 2n\pi,\\
\frac{\pi}{2} + 2n\pi &< \Omega_c\tau < \frac{3\pi}{2} + 2n\pi,
\label{Eq: omega_tau_limit}
\end{split}
\end{align}
where $n$ is an integer. They are used to find the limits of $\Omega_0, \Omega_c$ via the transcendental Eqs.~(\ref{Eq: trans_1}) and (\ref{Eq: trans_3}) respectively,
\begin{align}
\begin{split}
\omega - (k - 1)\epsilon &< \Omega_0 < \omega + (k - 1)\epsilon,\\
\omega - \epsilon &< \Omega_c < \omega + \epsilon.
\label{Eq: omega_limit}
\end{split}
\end{align}
The upper and lower limits of $\Omega_0$ and $\Omega_c$ are then used in Eq.~(\ref{Eq: omega_tau_limit}) to find the range of $\tau$ which are as follows, 
\begin{equation}
\frac{-\frac{\pi}{2} + 2n\pi}{\omega + (k - 1)\epsilon} < \tau < \frac{\frac{\pi}{2} + 2n\pi}{\omega - (k - 1)\epsilon},
\label{Eq: tau_inphase}
\end{equation}
for the in--phase solutions, 
\begin{equation}
\frac{-\frac{\pi}{2} + (2n + 1)\pi}{\omega + \epsilon} < \tau < \frac{\frac{\pi}{2} + (2n + 1)\pi}{\omega - \epsilon},
\label{Eq: tau_splay}
\end{equation}
for the splay phase solutions. We combine Eqs.~(\ref{Eq: tau_inphase}) and (\ref{Eq: tau_splay}) to write, 
\begin{equation}
\frac{(-\frac{1}{2} + n)\pi}{\omega + C\epsilon} < \tau < \frac{(\frac{1}{2} + n)\pi}{\omega - C\epsilon},
\label{Eq: tau_limit}
\end{equation}
where $C = k - 1$, $n$ is even for the in--phase branches and $C = 1$, $n$ is odd for the splay phase branches. The width of the $n$th $\tau$ interval, namely
\begin{equation}
I_n \in \left[ \frac{\pi(n - \frac{1}{2})}{\omega + C\epsilon}, \frac{\pi(n + \frac{1}{2})}{\omega - C\epsilon}\right]
\label{splay_sync_interval1}
\end{equation}
is 
\begin{equation}
L_n = \frac{2n\pi C\epsilon}{\omega^{2} - C^{2}\epsilon^{2}} + \frac{\pi\omega}{\omega^{2} + C^{2}\epsilon^{2}}.
\label{splay_sync_interval}
\end{equation}
The linear dependence of $L_n$ on $n$ implies that the solutions will overlap with increasing $\tau$, leading to multistability as shown in Fig. \ref{fig: freq_sync_splay}. For a given value of the delay parameter $\tau^{*}$, the total number of coexisting stable solutions can be determined graphically, as the total number of intersections of the branches with a vertical line at $\tau^{*}$ (see Fig.~\ref{fig: freq_sync_splay}). If $n$ is the index of the branch that intersects this vertical line at the highest value and $n - q$ the index of the  branch that intersects at the lowest, then there clearly are $q$ coexisting solutions. Branch $I_n$ overlaps with branch $I_{n - q}$ when the value of $\tau$ at the extreme right end of the latter exceeds the extreme left end of the former,  
\begin{equation}
\frac{n - \frac{1}{2}}{\omega + C\epsilon} < \frac{n - q + \frac{1}{2}}{\omega - C\epsilon}
\end{equation}
or 
\begin{equation}
q < \frac{\omega + 2nC\epsilon}{\omega + C\epsilon}, 
\end{equation}
giving 
\begin{equation}
q = \left\lfloor \frac{\omega + 2nC\epsilon}{\omega + C\epsilon} \right\rfloor,
\label{overlaps}
\end{equation}
where $\lfloor\cdot\rfloor$ is the floor function. Eq.~(\ref{overlaps}) gives the number of branches $q$ whose range of $\tau$ overlaps with the $\tau$ interval corresponding to $I_n$. Since $\tau = {n\pi}/{\omega}$ is within the interval $I_n$, the number of coexisting stable solutions (counting both the in- and splay phases) increases linearly with $\tau$ (see Fig. \ref{fig: freq_sync_splay_eps}a). The formula for coexisting solutions in Eq.~(\ref{overlaps}) further implies that for a given delay, $q$ increases linearly for small $\epsilon$, and levels off for higher coupling, as is evident in Fig. \ref{fig: freq_sync_splay_eps}b where the saturation behaviour of the number of stable solutions is evident. 
\section{\label{sec_3}Dimensional reduction}
\par In this section we consider the case of non-identical oscillators with different intrinsic frequencies as opposed the previous sections. Moreover  since we are considering ensembles of phase oscillators, it is interesting to examine the large $N$ limit through dimensional reduction, using the Ott-Antonsen technique \cite{OA_1, OA_2, OA_3}. Our intention is to either reduce the evolution equations to a single equation for each partition or to a single equation for the entire system. Following \cite{OA_3}, we briefly recall the necessary steps in the procedure here for this purpose. Define a complex order parameter for each oscillator as follows, 
\begin{equation}
R^{j}_{i}(t) = \sum\limits_{m = 1}^{k}\sum\limits_{l = 1}^{N_m} \frac{\a^{jm}_{il}}{K^{jm}_{i}}\exp(\iota\theta^{m}_{l}(t)),
\end{equation}
where as usual the superscript indexes the partitions and the subscript the oscillators within the partition, and rewrite Eq.~(\ref{Eq: evolve}) using this to get,
\begin{equation}
\dot{\theta^{j}_{i}} = \omega^{j}_{i} + \frac{-\iota\epsilon}{2}\left(R^{j}_{i}(t - \tau)\exp(-\iota\theta^{j}_{i})- 
R^{*j}_{i}(t - \tau)\exp(\iota\theta^{j}_{i})\right).
\label{Eq: evolve_new}
\end{equation}
\par Consider copies in which the connection topology is kept fixed (namely the same $\a$) but with the frequency of the $i$th oscillator in partition $j$ being chosen in the interval $[\omega^{\prime j}_{i} , \omega^{\prime j}_{i} + d\omega^{\prime j}_{i}]$ with probability $g(\omega^{\prime j}_{i})d\omega^{\prime j}_{i}$. The probability distribution of the intrinsic frequencies for a member of the ensemble is given by $\prod\limits_{j = 1}^{k}\prod\limits_{i =1}^{N_j}g(\omega^{j}_{i})d\omega^{j}_{i}$. The initial state of the system is specified by a probability distribution $f((\theta^{1}_{1},\omega^{1}_{1}),\ldots, (\theta^{k}_{N_k},\omega^{k}_{N_k}))$,which satisfies the evolution equation,
\begin{equation}
\frac{\partial f}{\partial t} + \sum\limits_{j = 1}^{k}\sum\limits_{i= 1}^{N_j} \frac{\partial [f\dot{\theta}^{j}_{i}]}{\partial \theta^{j}_{i}} = 0.
\label{Eq: conserve_1}
\end{equation}
Using the marginal probability distribution function $f^{j}_{i} = \int f\prod\limits_{(j, i) \neq (m, l)}d\omega^{m}_{l}d\theta^{m}_{l}$ one can rewrite Eq.~(\ref{Eq: evolve_new}) as
\begin{equation}
R^{j}_{i} = \sum\limits_{m = 1}^{k}\sum\limits_{l = 1}^{N_m}\frac{\a^{jm}_{il}}{K^{jm}_{i}}\int\limits_{-\infty}^{\infty}d\omega^{j}_{i}\int\limits_{0}^{2\pi}\exp(\iota\theta^{j}_{i}) f^{j}_{i}(\theta^{j}_{i}, \omega^{j}_{i}, t)d\theta^{j}_{i} 
\end{equation}
and it is straightforward to verify that $f^{j}_{i}$ satisfies the partial differential equation
\begin{equation}
\frac{\partial f^{j}_{i}}{\partial t} + \frac{\partial [f^{j}_{i} \dot{\theta}^{j}_{i}]}{\partial \theta^{j}_{i}} = 0.
\label{Eq: conserve_2}
\end{equation}
Using the Ott-Antonsen ansatz \cite{OA_3} one can find solutions of the above equations as
\begin{equation}
f^{j}_{i}(t) = \frac{g(\omega^{j}_{i})}{2\pi}\left[ 1 + \sum\limits_{n = 1}^{\infty}(p^{j}_{i}e^{i\theta^{j}_{i}})^{n} + \sum\limits_{n = 1}^{\infty}(p^{*j}_{i}e^{-i\theta^{j}_{i}})^{n}  \right],
\label{Eq: f_soln}
\end{equation}
with  $|p^{j}_{i}| < 1$ to ensure convergence of the solutions. When the frequencies  $g(\omega^{j}_{i})$ are taken from a Lorentzian distribution, namely $$g(\omega) = \Delta/(\pi[(\omega - \omega_0)^2 + \Delta^2]),$$ where $\omega_0$ and $\Delta$ are the location and the scale parameters, then following \cite{OA_3}, the solution of $f^{j}_{i}(t)$ in Eq.~(\ref{Eq: f_soln}) will satisfy Eq.~(\ref{Eq: conserve_2}) if
\begin{align}
\begin{split}
\frac{dp^{j}_{i}(t)}{dt} &+ \frac{\epsilon R^{j}_{i}(t - \tau)}{2}\left[ (p^{j}_{i}(t))^2 - 1 \right] + \Delta p^{j}_{i}(t) = 0,\\
R^{j}_{i}(t) &= \sum\limits_{m = 1}^{k}\sum\limits_{l = 1}^{N_m}\frac{\a^{jm}_{il}}{K^{jm}_{i}}p^{m}_{l}(t),\\
\label{Eq: evolve_p_r}
\end{split}
\end{align}
where $R^{j}_{i}$ and $p^{j}_{i}$ are real. 

So far there is no dimensional reduction since the number of equations specified by Eq.~(\ref{Eq: evolve_p_r}) are the same as the original flow equations, Eq.~(\ref{Eq: evolve}). However if there is a uniform degree distribution at each node, then by averaging Eq.~(\ref{Eq: evolve_p_r}) over initial conditions specified by $f$ at $t = 0$, one can reduce the dimension. For the case of $p^{j}_{i}(t) = p(t)$ for all $j$ and $i$, a special solution can be obtained since Eq.~(\ref{Eq: evolve_p_r}) becomes,
\begin{equation}
\frac{dp(t)}{dt} + \frac{\epsilon(k - 1)p(t - \tau)}{2}[p^{2}(t) - 1] + \Delta p(t) = 0,
\label{Eq: reduced_OA}
\end{equation}
with $R^{j}_{i} = (k - 1)p(t - \tau)$.
\begin{figure}
\begin{tabular}{cc}
    \includegraphics[scale = 0.5]{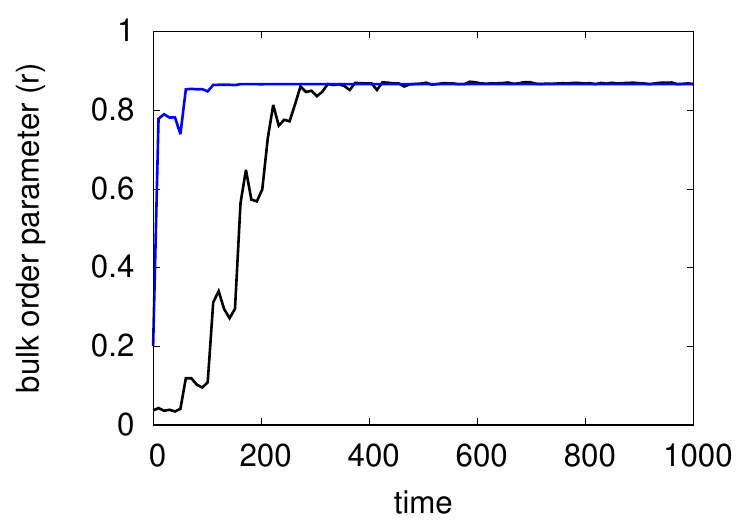}&
    \includegraphics[scale = 0.68]{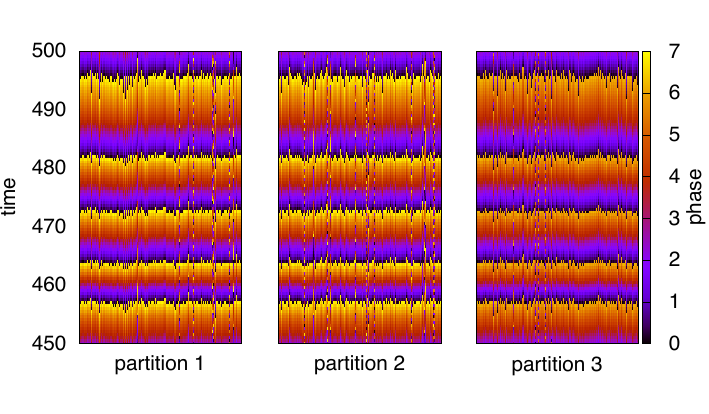}\\
    (a) & (b)\\
\end{tabular}
\caption{\footnotesize (a) Variation of the bulk order parameter calculated via Eq.~(\ref{Eq: reduced_OA}) using Ott-Antonsen ansatz (in blue). Shown in black are the results obtained for the bulk order parameter by direct simulation of a tripartite ensemble with $N$= 600 nodes. Other parameters are $\epsilon = 0.8, \tau = 50$. (b) Space-time plot of the phases of the system obtained by evolving Eq.~(\ref{Eq: evolve}) for the same set of parameters. There are 200 oscillators in each partition and their intrinsic frequencies are chosen from a Lorentzian distribution with $\Delta = 0.2$ and $\omega_0 = 0.6$. The degree of each node in the system is chosen as 50 and each is randomly connected to the nodes of the other partitions with the condition that there exists at least one link to each of the other partitions.}
\label{fig: OA}
\end{figure}

The utility of the dimensional reduction can be seen by comparing with numerical simulations. Shown in Fig.~\ref{fig: OA}  is the bulk order parameter, $r = \frac{1}{N(k - 1)}\sum\limits_{j = 1}^{k}\sum\limits_{i=1}^{N_j}R^{j}_{i}$, for the special case of {\em equal} degree distribution. In the steady state, the bulk order parameter obtained from numerically simulating Eq.~(\ref{Eq: evolve}) or from the above analysis are identical, indicating the applicability of the OA ansatz to deduce the asymptotic nature of the dynamics in a multipartite system of arbitrary size and number of partitions.  When the intrinsic frequencies of the oscillators are drawn from a distribution, then in the thermodynamic limit the system is attracted to a manifold given by the coefficients, $p$. The steady state is partially synchronised, with a small number of oscillators out of synchrony (see the space-time plot of the steady state dynamics in Fig. \ref{fig: OA}.(b)). Our results here are in conformity with the findings of Pikovsky and Rosenblum \cite{pikovsky2008} and Martens \cite{martens2010} for limit cycle oscillators with phase shifts in coupling, as well the work on emergent behaviour in modular networks \cite{ujjwal2016}. In multipartite systems when all oscillators have identical frequencies or when the frequencies within partitions are identical (and possibly distinct from partition to partition) there can be multiple coexisting attractors where the system is phase locked. The final steady state then depends on the initial conditions and the present analysis does not carry over. 

\section{\label{sec_4}Summary}
\par In this work our aim has been to examine the nature of the collective dynamics in networks with a multipartite topology. Motivation for studying such networks arise from the fact that any given graph of size $N$ could be viewed as a $k$-partite network for some $1\le k \le N$. In earlier work we have studied the bipartite case, $k$ = 2 extensively \cite{punetha2015, atay2015}. The limiting cases, $k$ = N and $k = 1$ are achieved, respectively, when the nodes are all unconnected and when every node is connected to every other node. We therefore consider networks of time-delay coupled phase oscillators on a multipartite network consisting of $k$ partitions, with $2 < k < N$. 

\par In the multipartite networks considered, the number of nodes in each of the partitions can vary, and furthermore, the coupling between the partitions can also be quite general. We are therefore able to extend earlier studies of phase locked dynamics in time-delayed bipartite networks \cite{schuster1989,earl2003, punetha2015, atay2015}.  The present approach makes it possible to easily determine the dynamics and stability of networks of complex architecture.   The present results can be effectively applied to find steady state solutions on complex networks that are  multipartite, but with the requirement that each node has at least one link to each of the other partitions. We believe that our results should also apply to similar systems with phase lags in the coupling since the lag parameter can be interpreted as a discrete time-delay for small $\tau$. 

\par Numerical results have been presented mainly for the tripartite case, $k$=3. When the initial frequencies of all oscillators are identical, then for sufficiently large coupling, the ensemble of phase oscillators synchronize in phase in a manner similar to that observed in the Kuramoto model. In addition there is the stable splay cluster solution where each partition is in perfect synchrony, but between partitions there is a phase shift of $2\pi/k$ as seen in delay-coupled limit cycle oscillators \cite{atay2015} for the case of $k$=2. Other phase locked solutions are also observed, but these are dependent on the parameters as well as the number of partitions of the system.  A generalised stability criteria for the phase-locked solutions for the case of $k>2$ is given, extending the results for the bipartite network. One of the striking features of such systems is the existence of a large number of coexisting stable states; this number grows linearly with the time delay and varies sigmoidally with the coupling strength. 

\par It has also been possible to obtain a low-dimensional representation of the extended system through the use of the Ott-Antonsen ansatz \cite{OA_1, OA_3}. In the order parameter formulation of the system and for a unimodal distribution of frequencies, the multipartite system is described by a single evolution equation in the case when the nodes have identical degrees.  Under this restriction, we show that the dynamical equation describes the asymptotic evolution of the system onto a  manifold on which the system is partially synchronised. 

\par
\begin{figure}[H]
\centering
\includegraphics[scale = 0.5]{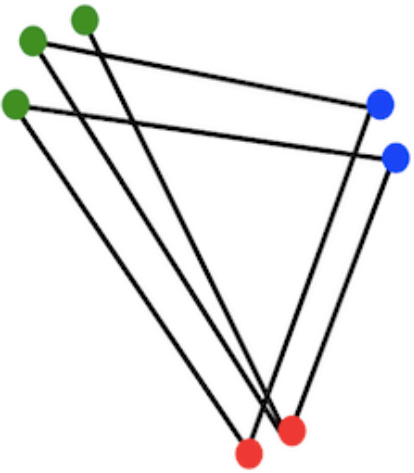}
\caption{\footnotesize A tripartite network in which a node in one of the partitions (in green)  is not linked to one of the other two partitions (in blue). 
Such a network does not support splay states, although the fully synchronized solution is stable.}
\label{fig: diff_topology}
\end{figure}

 An important caveat needs to be mentioned for networks with $k>$ 2. The existence of the generic in-phase and splay phase solutions depends on the existence of nonzero {\em partial degrees} of each node. In case of a null partial degree (see Fig. \ref{fig: diff_topology}), the equations involving $\Omega, \tau$ and $\mu$ are obtained in addition to Eq.~(\ref{Eq: OMG}). For example, in the case of a tripartite network, if there is a node in one partition that is not connected to at least one node in each of the other partitions, one gets the additional condition, $\sin(- \Omega\tau - \mu_m + \mu_j) = 0$. With $M$ such nodes, there will be $M$ such equations.  For $k > 3$, $\sum\limits_{j} \sin(- \Omega\tau -\mu_m + \mu_j) = 0,$ where the summation is over all the partitions to which links are missing for a node. Clearly symmetry in Eq.~(\ref{Eq: OMG}) which existed for $\mu_j = 0$ and $\bar{\mu}_{jm} = {2\pi}/{k}$ will be destroyed, and the above result for splay states then does not hold. 

\par The nature of the basins of attraction of each of the phase-locked solutions and their dependence on the network structure are additional features of interest. What role can be played by varying the nature of the interactions -- having both attractive and repulsive forms -- and having a distribution in the time-delays are also questions that need to be explored in future work.

\section*{Acknowledgement}
We thank Nirmal Punetha for extensive discussions during the course of this work. RR is a recipient of the J. C. Bose National Fellowship of the Science and Engineering Research Board, India. JS acknowledges the support of Indian Institute of Technology Delhi, India, in the form of Institute Post Doctoral Fellowship (IPDF).   

\appendix
\section{\label{app: sec_1}Stability of the phase locked states}
\par Consider the $k-$partite system with a normalised adjacency matrix, $\widetilde{\a}$, given in terms of the normalised blocks with elements, $\a^{jm}_{il}/K^{jm}_{i}$ and a general coupling function $h$ which is $2\pi$-periodic in its argument and differentiable up to first order. We rewrite Eq.~(\ref{Eq: evolve}) as, 
\begin{equation}
    \dot{\theta}^{j}_{i}(t) = \omega_{j} + \sum_{m=1}^{k}\sum_{\ell=1}^{N_m}\epsilon\widetilde{\a}^{jm}_{i\ell}h(\theta^{m}_{\ell}(t - \tau) - \theta^{j}_{i}(t)).
    \label{Eq: evolve_app}
\end{equation}
To analyse the stability we perturb the general phase locked state of the above evolution equation,
\begin{equation}
    \theta^{j}_{i} = \Omega t - \mu_{j} + \xi^{j}_{i}(t),
    \label{Eq: pert_soln}
\end{equation}
where $j = 1, 2, \cdots, k$. Using Eq.~(\ref{Eq: evolve_app}) one can see that the evolution equations for the perturbations are 
\begin{equation}
        \dot{\xi}^{j}_{i}(t) = \sum\limits_{m = 1, m \neq j}^{k} h'(-\Omega\tau -\mu_m + \mu_j) F_m(\xi^{j}_{i}),       
         \label{Eq: pert}
    \end{equation}
where  $F_m(\xi^{j}_{i}) = \sum\limits_{l = 1}^{N_m}\frac{\epsilon \a^{jm}_{il}}{K^{jm}_{i}}(\xi^{m}_{l}(t - \tau) - \xi^{j}_{i}(t))$. The stability of the synchronized solution and the general form of the clustered solutions can now be analysed.\\
\subsection{Stability of the in-phase solutions (\texorpdfstring{$\mu_j = 0$}{})}
\par {\bf Proposition I} : The in-phase solution of the $k-$partite system is stable if and only if $h'(-\Omega\tau) > 0$. 
\par \textit{Proof} : For the in-phase solutions, Eq.~(\ref{Eq: pert}) becomes, 
\begin{equation}
    \dot{\xi}^{{j}}_{i}(t) = h'(-\Omega\tau)\sum\limits_{m = 1, m \neq j}^{k}F_m(\xi^{j}_{i}).
\end{equation}
Writing  $\xi^{j}_{i}(t) = v^{j}_{i}\exp(\lambda t)$ and using the definition of $F_m$, it is straightforward to obtain (for $j = 1, 2, \ldots, k$)
\begin{equation}
    \frac{\exp(\lambda \tau)\left[ \lambda + (k - 1)\epsilon h'(-\Omega \tau)\right]v_{i}^{j}}{\epsilon h'(-\Omega \tau)} 
    = \sum\limits_{m = 1}^{k} \sum\limits_{l = 1}^{N_m}\widetilde{\a}^{jm}_{il}v_{l}^{m}.
    \label{Eq: eig_pert}
\end{equation}

Clearly this relationship can be cast as an eigenvalue equation, 
\begin{equation}
\mathbf{M} v= \zeta v.
\end{equation}\noindent
with appropriate $\mathbf{M}$ and eigenvalue $\zeta = \dfrac{\exp(\lambda \tau)[\lambda + (k - 1)\epsilon h'(-\Omega\tau)]}{\epsilon h'(-\Omega\tau)}$. We rewrite $\zeta = |\zeta|\exp(i\Phi)$,  $\lambda = r + i s$  and separating real and imaginary parts gives the pair of equations
\begin{align}
    \begin{split}
        \epsilon h'(-\Omega\tau)|\zeta|\cos(\Phi - s\tau)e^{-r\tau} &= r + ((k - 1)\epsilon h'(-\Omega\tau)),\\
        \epsilon h'(-\Omega\tau)|\zeta|\sin(\Phi - s\tau)e^{-r\tau} &= s,
    \end{split}
    \label{Eq: r_1}
\end{align}
from which one obtains,
\begin{equation}
|\zeta|^2 e^{-2r\tau} = (k - 1)^2 + \frac{r^2 + s^2 + 2r(k - 1)\epsilon h'(-\Omega\tau)}{|\epsilon h'(-\Omega\tau)|^{2}}.
\label{Eq: in_phase}
\end{equation}
Our objective is to find the relationship between the signs of $r$ and $\epsilon h'(-\Omega\tau)$. The Gershgorin theorem \cite{berkey1975} can be used to estimate the upper bound of eigenvalue, $\zeta$ of the matrix $\mathbf{M}$. In the complex plane of eigenvalues of $\mathbf{M}$, each Gershgorin disk has its centre at the origin since the diagonal entries of the matrix are zero. As the blocks of $\mathbf{M}$ are given by $\a^{jm}_{il}/K^{jm}_{i}$, the radii of the Gershgorin disks are given as $k - 1$, giving the upper bound of $|\zeta|$ as $k - 1$. This implies that for $r = |r| > 0$, the following inequality, $0 < |\zeta|^{2}e^{-2|r|\tau} < (k - 1)^{2}$ must always be satisfied. On the other hand if both, $\epsilon h'(-\Omega\tau) > 0$ and $r > 0$, then the R.H.S. of Eq.~(\ref{Eq: in_phase}) is greater than $(k - 1)^2$. This implies that when $r = |r| > 0$, the quantity $\epsilon h'(-\Omega\tau)$ cannot be simultaneously positive.

\par We now show that when $r$ is positive, the sign of $\epsilon h'(-\Omega\tau)$ is in fact negative. Rearranging the first relation in Eq. \ref{Eq: r_1} and using $r = |r| > 0$, we obtain, 
\begin{equation}
\epsilon h'(-\Omega\tau) = \frac{|r|}{\rho e^{-|r|\tau} - (k - 1)},
\label{Eq: r_6}
\end{equation}
where $\rho = |\zeta|\cos(\phi - s\tau)$. It is clear that the following inequality, $-(k - 1) < \rho < k - 1$ or $ -(k - 1)e^{-|r|\tau} < \rho e^{-|r|\tau} < (k - 1)e^{-|r|\tau} < k - 1$, is always true. This indicates that the denominator, $\rho e^{-|r|\tau} - (k - 1)$ in Eq. \ref{Eq: r_6} is negative which implies $\epsilon h'(-\Omega\tau) < 0$.  
\par
\begin{figure}[H]
\centering
\includegraphics[scale = 0.5]{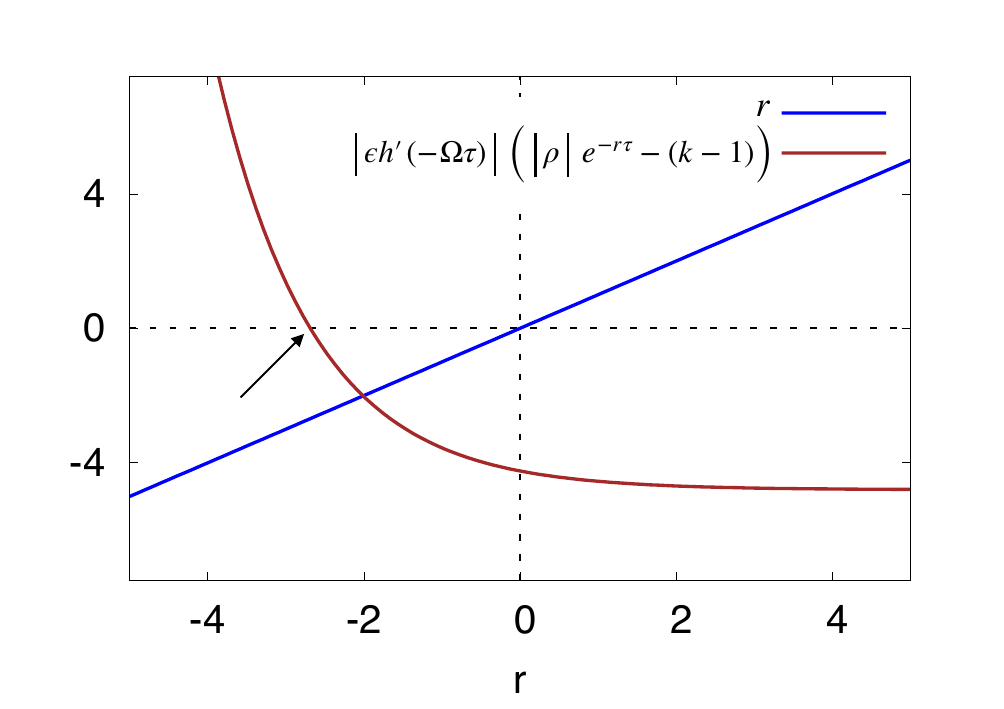}
\caption{\footnotesize A graph of $r$ and $|\epsilon h'(-\Omega\tau)|(|\rho|e^{-r\tau} - (k - 1))$ plotted against $r$ for $\epsilon h'(-\Omega\tau) = 0.8$, $\rho = 0.7$, $\tau = 0.8$ and $k = 7$. The arrow denotes the point $r = \frac{1}{\tau}\log_{e}\left(\frac{|\rho|}{k - 1}\right)$ where the function $|\epsilon h'(-\Omega\tau)|(|\rho|e^{-r\tau} - (k - 1))$ becomes negative for all $r > \frac{1}{\tau}\log_{e}\left(\frac{|\rho|}{k - 1}\right)$.}
\label{fig: negative_r}
\end{figure}

 We now show the sufficiency of the stability criteria. For $\epsilon h'(-\Omega\tau) = |\epsilon h'(-\Omega\tau)|> 0$, the first relationship in Eq.~(\ref{Eq: r_1}) becomes, 
\begin{equation}
    |\epsilon h'(-\Omega\tau)|\rho e^{-r\tau} = r + (k - 1)|\epsilon h'(-\Omega\tau)|,
    \label{Eq: r_2}
\end{equation}
where, $-(k - 1) < \rho = |\zeta|\cos(\Phi - s\tau) < k - 1$. When $\rho = |\rho| > 0$, Eq.~(\ref{Eq: r_2}) becomes 
\begin{equation}
r = |\epsilon h'(-\Omega\tau)|(|\rho|e^{-r\tau} - (k - 1)).
\label{Eq: r_4} 
\end{equation}
We argue that the above equation always has a solution at a negative value of $r$. The R.H.S. of the above equation is always positive for $r < \frac{1}{\tau}\log_{e}\left(\frac{|\rho|}{k - 1}\right)$ and negative for $r > \frac{1}{\tau}\log_{e}\left(\frac{|\rho|}{k - 1}\right)$. Since we have already established that $\rho < k - 1$, the sign of the point $r = \frac{1}{\tau}\log_{e}\left(\frac{|\rho|}{k - 1}\right)$ is negative. This implies that the L.H.S. and R.H.S. of Eq.\ref{Eq: r_4} will always intersect each other in the third quadrant for a negative value of $r$. This can also be visualised by plotting the left and the right hand sides of Eq.~(\ref{Eq: r_4}) with respect to $r$ in Fig.~\ref{fig: negative_r} for a typical set of values of $\epsilon h'(-\Omega\tau)$, $\rho$, $k$ and $\tau$. We thus see from Fig.~\ref{fig: negative_r} that there exists a solution to Eq.~(\ref{Eq: r_4}) with $r < 0$.
For $\rho = -|\rho| < 0$, one obtains $r = -(k - 1)|\epsilon h'(-\Omega\tau)| - |\epsilon h'(-\Omega\tau)|\rho|e^{-r\tau}$, which is always negative. Therefore whenever $h'(-\Omega\tau) > 0$, $r$ is negative, verifying that the in-phase solution is stable. Thus the sufficiency of the stability criteria of the in-phase solution is proved.

\subsection{Stability criteria for the $k-$cluster solution (\texorpdfstring{$\mu_j \neq 0$}{})}
\par {\bf Proposition II} : A $k$-cluster solution of the $k-$partite system is stable iff $\forall j = 1, 2, \ldots, k$,  $S_j = 
\sum\limits_{m = 1, m \neq j}^{k}\epsilon h'(-\Omega \tau - \mu_m + \mu_j) > 0$.

\textit{Proof}: The perturbation $\xi^{j}_{i}(t)$ to the $i$th phase oscillator in partition $j$ evolve according to Eq.~(\ref{Eq: pert}), and using $\xi^{j}_{i} = v^{j}_{i}\exp(\lambda t), j = 1, 2, \cdots, k$, Eq.~(\ref{Eq: pert}) can be is written as, 
\begin{equation}
(\lambda + S_j)e^{\lambda\tau}v^{j}_{i} = \sum\limits_{m = 1, m \neq j}^{k}\sum\limits_{l = 1}^{N_m}\frac{\epsilon\a^{jm}_{il}}{K^{jm}_{i}}h'(-\Omega\tau - \mu_m + \mu_j)v^{m}_{l}.
\label{Eq: eig_prop_II}
\end{equation}
The above relation can be rewritten as an eigenvalue equation, $\widetilde{\mathbf M}v = \exp(\lambda \t) v.$ The blocks of $\widetilde{\mathbf{M}}$ are written in terms of the blocks of the matrix $\widetilde{\a}$ as follows, 
\begin{equation}
        \widetilde{\mathbf{M}}^{jm} = \frac{\widetilde{\a}^{jm}\epsilon h'(-\Omega\tau - \mu_m + \mu_j)}{\lambda + S_j}.
\end{equation}
Again, by the Gershgorin theorem \cite{berkey1975}, the upper bound of the eigenvalues of  $\widetilde{\mathbf{M}}$ is given by the maximum of $U_j = \dfrac{S^{+}_{j}}{|\lambda + S_{j}|}$, $j = 1, 2, \cdots, k$, where $S^{+}_{j} = \sum\limits_{m = 1, m \neq j}^{k}|\epsilon h'(-\Omega\tau - \mu_m + \mu_j)|$ i.e. the sum of the absolute values of each term in $S_j$. Writing $\lambda = r + i s$, it is clear that the absolute value of eigenvalue of $\widetilde{\mathbf{M}}$, $|\exp(\lambda\tau)| < 1$ (resp. $>1$) when $r < 0$ (resp. $>0$). Therefore when $r < 0$, $U_j, j = 1, 2, \cdots, k$ can be either greater than one or less than one while for $r > 0$, at least one of $U_j$s should be greater than one. Therefore the relationship between the signs of $r$ and $S_j$'s should be in conformity with both of the conditions, $U_j < 1$ and $U_j > 1$.  

In the case $U_j > 1$, using $\lambda = r + \iota s$ we get the inequality
\begin{align}
    \begin{split}
     (S^{+}_{j})^2 - (S_{j})^2 &> |\lambda|^2 + 2rS_j.
    \end{split}
\end{align}
The left hand side of the above inequality is either a positive quantity, $c_0$ or zero. Therefore any of the following conditions are possible : $c_0 > |\lambda|^2 + 2rS_j = ||\lambda|^2 + 2rS_j| > 0$, $|\lambda|^2 + 2rS_j = -||\lambda|^2 + 2rS_j| < 0$ and a third possibility, $|\lambda|^2 + 2rS_j = 0$.
\par To prove the proposition for the first possibility, $c_0 > |\lambda|^2 + 2rS_j > 0$, we first show that when $r < 0$, $S_j > 0$. Using $r = -|r|$, we get,
\begin{equation}
c_0 > |\lambda|^2 - 2|r|S_j > 0,
\end{equation}
which can be rearranged as follows,
\begin{equation}
\frac{c_0 + 2|r|S_j}{S_j} > \frac{|\lambda|^2}{S_j} > 2|r|.
\end{equation}
Since $2|r|$ is a positive quantity, for the above inequality to be true, we must have $S_j = |S_j|$. The condition is also sufficient i.e. $r > 0$ $\forall$ $S_j < 0$; this can be seen by noting that,
\begin{equation}
c_0 > |\lambda|^2 - 2r\left|S_j\right| > 0.
\end{equation}
We can rearrange the above inequality as,
\begin{equation}
\frac{c_0 + 2r\left|S_j\right|}{r} > \frac{|\lambda|^2}{r} > 2\left|S_j\right|.
\end{equation}
Following the similar argument as before, since $2|S_j| > 0$, the above inequality is true only when $r > 0$. 

\par Next we prove the proposition for the second possibility, $|\lambda|^2 + 2rS_j < 0$. Replacing $r$ by $-|r|$ gives, $|\lambda|^2 - 2|r|S_j < 0$ or $\dfrac{|\lambda|^2}{2|r|} < S_j$. Since all of the quantities on the left are positive, $S_j$ is always positive. After a little algebra, the sufficiency of the criteria, $r = |r| > 0$ $\forall$ $S_j = -|S_j| < 0$, can also be shown. In this case we have, $|\lambda|^2 - 2r|S_j| < 0$ or $\dfrac{|\lambda|^2}{2|S_j|} < r$, implying that $r$ is positive. 

\par The relationship between the signs of $r$ and $S_j$ can be easily verified for the third case, $|\lambda|^2 + 2rS_j = 0$. We observe that when $r = -|r| < 0$, $S_j = \dfrac{|\lambda|^2}{2|r|}$ and $r = \dfrac{|\lambda|^2}{2\left|S_j\right|}$ $\forall$ $S_j = -\left| S_j\right| < 0$. Thus the proof of our proposition for the stability criteria is complete for the case $U_j > 1$.

\par In the case of $U_j < 1$, similar analysis as before gives,
\begin{align}
    \begin{split}
        (S^{+}_{j})^2 - (S_{j})^2 &< |\lambda|^2 + 2rS_j.
    \end{split}
\end{align}
The above inequality implies two possibilities, $|\lambda|^2 + 2rS_j > 0$ and $|\lambda|^2 + 2rS_j > c_0$ with $c_0 > 0$. In the first case, one can easily observe that when $r < 0$, $S_j  > 0$. Using $r = -|r|$, we get $|\lambda|^2 - 2|r|S_j > 0$ or $\dfrac{1}{S_j} > \dfrac{2|r|}{|\lambda|^2}$ which implies that $S_j$ is positive since the right hand term is always positive. In this case also,  $r > 0$ $\forall$ $S_j < 0$; this can be seen by noting that $S_j = -|S_j|$ and therefore $|\lambda|^2 - 2r|S_j| > 0$ or $\dfrac{1}{r} > \dfrac{2|S_j|}{|\lambda|^2}$. This indicates that $r$ is positive since the right hand term in the last inequality is always positive. Thus the sufficiency of the proof of the correlation between the signs of $r$ and $S_j$ is shown. 

\par For the second case, we first use $r = -|r| < 0$ and get $|\lambda|^2 - 2|r|S_j > c_0$. As we have assumed that $c_0 > 0$, it must be true that $|\lambda|^2 > 2|r|S_j$ or $\dfrac{1}{S_j} > \dfrac{2|r|}{|\lambda|^2}$. Therefore $S_j$ must be positive for all $r < 0$. We can similarly argue the sufficiency of this proof as follows : for $S_j = -\left| S_j\right|$, we have $|\lambda|^2 - 2r\left|S_j\right| > c_0$. As $c_0 > 0$, we must have, $|\lambda|^2 > 2r\left|S_j\right|$ or $\dfrac{1}{r} > \dfrac{2\left|S_j\right|}{|\lambda|^2}$. 

\par Thus we complete the proof of our proposed stability criteria. Additionally one can see that when $\mu_{j} = 0$, the stability criteria given in Proposition II reduces to the criteria given in Proposition I.  

\nocite{*}

\bibliography{reference}

\end{document}